\documentclass{emulateapj}
\usepackage{apjfonts,lscape}

\shorttitle{\sc A cosmic abundance standard}
\shortauthors{\sc Przybilla et al.}

\begin{document}

\title{
A cosmic abundance standard: chemical homogeneity of the solar neighbourhood\\ 
and the ISM dust-phase composition
\altaffilmark{1}
}


\author{Norbert Przybilla\altaffilmark{2}, Maria-Fernanda
Nieva\altaffilmark{2,3} 
and Keith Butler\altaffilmark{4}}
\email{przybilla@sternwarte.uni-erlangen.de}
\altaffiltext{1}{Based on observations obtained at the European Southern
Observatory, proposal 074.B-0455(A).}
\altaffiltext{2}{Dr. Remeis-Observatory, 
Sternwartstr.\,7, D-96049 Bamberg, Germany}
\altaffiltext{3}{MPI for Astrophysics,
Postfach 1317, D-85741 Garching, Germany}
\altaffiltext{4}{University Observatory, Scheinerstr. 1, D-86179 Munich, Germany}




\begin{abstract}
A representative sample of unevolved early B-type stars in nearby OB associations 
and the field is analysed to unprecedented precision using NLTE techniques. 
The resulting chemical composition is found to be more metal-rich and much more 
homogeneous than indicated by previous work. A rms scatter of $\sim$10\% in 
abundances is found for the six stars (and confirmed by six
evolved stars), the same as reported for ISM gas-phase abundances. 
A cosmic abundance standard 
for the present-day solar neighbourhood
is proposed, implying mass fractions for hydrogen, helium and
metals of $X$\,$=$\,0.715, $Y$\,$=$\,0.271 and $Z$\,$=$\,0.014. Good
agreement with solar photospheric abundances as reported from recent 3D 
radiative-hydrodynamical simulations of the solar atmosphere is obtained. 
As a first application we use the cosmic abundance standard as a proxy for the
determination of the local ISM dust-phase composition, 
putting tight observational constraints on dust models.
\end{abstract}






\section{Introduction}\label{intro}
The Sun is unique among the stars because independent~in\-dicators allow its
chemical composition to be constrained with a precision unmatched for any
other star. This can
be done by spectroscopic analysis of its photosphere and by 
measurement of solar wind and solar energetic particles. Solar nebula
abundances can be determined from CI 
chon\-drites, which are unaltered since the formation of the system.
The wealth of information established the Sun~as~the principal
standard for the chemical composition of cosmic matter 
(e.g. Grevesse \& Sauval 1998, GS98; Holweger 2001; Asplund et al 2005,
AGS05). 
However, is a 4.6\,Gyr old star indeed representative of the cosmic matter in its
neighbourhood\footnote{
We consider the region
at distances shorter than $\sim$1\,kpc (and $\pm$500 pc in
Galactocentric direction) 
as solar neighbourhood in order to minimize bias
due to Galactic abundance gradients, see Fig.~\ref{evol} for a schematic
overview.} 
at present?

Ideal indicators for pristine abundances
are unevolved early B-stars of spectral types B0--B2. 
Slowly rotating stars are preferred as
their photospheres should be essentially unaffected by mixing of
CN-processed material 
\citep[][]{maeder00}. The
atmospheres of early B-stars are also unaffected by atomic diffusion
that gives rise to peculiarities of metal abundances in many
later-type stars \citep[e.g.][]{smith96}. 
A major practical advantage is also their relatively simple photospheric
physics, which is represented well by 
classical model atmospheres,
unaffected by complications such as stellar winds
or convection.

As a consequence, early B-stars in the solar neighbourhood were subject of
several NLTE studies in the past
\citep[e.g.][]{gies92,kilian92,kilian94,cunha94,daflon99,daflon01a,daflon01b,daflon03,lyubimkov04,lyubimkov05}.
Overall, they found a wide range of abundances, by about a factor $\sim$10,
and an average metallicity of only $\sim$2/3 solar (GS98). 
Hence, the impression arose that the solar neighbourhood is chemically highly
heterogeneous, and the Sun anomalously metal-rich compared to young stars.

Both findings are problematic in
terms of Galactic chemical evolution. Dispersal of stellar nucleosynthesis
products increases the metallicity 
over time \citep[e.g.][]{chiappini03} and hydrodynamic mixing tends to
homogenize the interstellar medium (ISM) 
locally \citep{edmunds75}. Characteristic timescales
for homogenization are short, ranging from 10$^6$--10$^8$\,yrs on scales of
100-1000\,pc \citep{roy95}. 

In contrast to the young stars the interstellar gas shows a high 
degree of chemical homogeneity in the solar neighbourhood \citep{sofia04},
with the 
rms scatter of mean abundances being $\sim$10\%. 
However, the ISM gas phase is not suitable as a tracer for cosmic
abundances because of selective depletion
of elements onto dust grains. 
Here we reinvestigate the conundrum of inhomogeneous stellar vs. homogeneous
ISM gas-phase abundances in the solar neighbourhood, motivated by the
finding of homogeneous B-star abundances for carbon
\citep[see detailed analysis by][NP08]{nieva08}. 


\section{Sample Analysis}
Six bright and apparently slow-rotating early B stars in the solar
neighbourhood -- randomly distributed in
OB associations and in the field, and covering a wide range of stellar parameters -- 
were observed in early 2005 at ESO/La Silla,
using FEROS on the 2.2\,m telescope.
Spectra with broad wavelength coverage and resolving power $\lambda/\Delta
\lambda$\,$\approx$\,48\,000 were
obtained, at very high-S/N (up to $\sim$800 in the $B$-band).

{
\begin{table*}
\footnotesize
\caption{Stellar parameters \& elemental abundances\label{params}}
\setlength{\tabcolsep}{1.5mm}
\begin{center}
\vspace{-5mm}
\begin{tabular}{lrrrrrr}
\tableline\\[-2.5mm]
\tableline
                            & \object[HR 6165]{HR\,6165} & \object[HR 3055]{HR\,3055} & \object[HR 1861]{HR\,1861} & \object[HR 2928]{HR\,2928} & \object[HR 3468]{HR\,3468} & \object[HR 5285]{HR\,5285}\\
\tableline
Sp. Type                    &              B0.2\,V &              B0\,III &               B1\,IV &               B1\,IV &            B1.5\,III &                B2\,V\\
Association                 &              Sco Cen &                Field &             Ori OB1b &                Field &                Field &              Sco Cen\\[.5mm]
$d$\,(pc)                   &           152$\pm$20 &           438$\pm$57 &           450$\pm$59 &            481$\pm$63 &           319$\pm$41 &            155$\pm$20 \\
$T_{\rm eff}$\,(K)          &        32000$\pm$300 &        31200$\pm$300 &        27000$\pm$300 &        26300$\pm$300 &        22900$\pm$300 &  20800$\pm$300\\
$\log g$\,(cgs)             &        4.30$\pm$0.05 &        3.95$\pm$0.05 &        4.12$\pm$0.05 &        4.15$\pm$0.05 &        3.60$\pm$0.05 &        4.22$\pm$0.05\\
$\xi$\,(km\,s$^{-1}$)       &              5$\pm$1 &              8$\pm$1 &              3$\pm$1 &              3$\pm$1 &              5$\pm$1 &              3$\pm$1\\
$v\sin i$\,(km\,s$^{-1}$)   &              4$\pm$4 &             29$\pm$4 &             12$\pm$1 &             14$\pm$1 &             11$\pm$2 &             18$\pm$1\\
$\zeta$\,(km\,s$^{-1}$)     &              4$\pm$4 &             37$\pm$8 &              \nodata &             20$\pm$2 &             20$\pm$1 &              \nodata\\
\tableline
$\varepsilon$(He)$^{\rm a}$           &  10.99$\pm$0.05~(20) &  10.94$\pm$0.05~(16) &  10.99$\pm$0.05~(14) &  10.99$\pm$0.05~(14) &  10.99$\pm$0.05~(14) &  10.99$\pm$0.05~(13)\\
$\varepsilon$(C\,{\sc ii})$^{\rm b}$  &   8.27$\pm$0.14~(13) &   8.35$\pm$0.08~(10) &   8.32$\pm$0.10~(19) &   8.28$\pm$0.08~(18) &   8.36$\pm$0.10~(17) &   8.32$\pm$0.08~(20)\\
$\varepsilon$(C\,{\sc iii})$^{\rm b}$ &   8.31$\pm$0.11~(17) &  8.30$\pm$0.05~~~(7) &   8.36$\pm$0.03~(11) &   8.27$\pm$0.02~~~(5) &   8.47$\pm$0.04~~~(2) &   8.42$\pm$0.06~~~(2) \\
$\varepsilon$(C\,{\sc iv})$^{\rm b}$  &           8.34~~~(2) &           8.45~~~(2) &   \nodata & \nodata & \nodata & \nodata\\
$\varepsilon$(N\,{\sc ii})$^{\rm c}$  &   8.16$\pm$0.12~(73) &   7.77$\pm$0.08~(23) &   7.75$\pm$0.09~(61) &   8.00$\pm$0.12~(61) &   7.92$\pm$0.10~(56) &   7.76$\pm$0.08~(47) \\
$\varepsilon$(O\,{\sc i})$^{\rm d}$   &              \nodata &              \nodata &  8.82$\pm$0.03~~~(3) & 8.83$\pm$0.05~~~(5) & 8.82$\pm$0.03~~~(7) & 8.79$\pm$0.05~~~(7) \\
$\varepsilon$(O\,{\sc ii})$^{\rm e}$  &   8.77$\pm$0.08~(51) &   8.79$\pm$0.10~(41) &   8.74$\pm$0.11~(52) &   8.74$\pm$0.09~(46) &   8.80$\pm$0.09~(40) &   8.71$\pm$0.05~(45) \\
$\varepsilon$(Ne\,{\sc i})$^{\rm f}$  &  8.12$\pm$0.05~~~(2) &              \nodata &  8.12$\pm$0.08~~~(9) & 8.11$\pm$0.09~~~(9) & 8.05$\pm$0.09~(10) & 8.07$\pm$0.07~(14)\\
$\varepsilon$(Ne\,{\sc ii})$^{\rm f}$ &   8.14$\pm$0.07~(16) &  8.07$\pm$0.07~~~(8) &   8.08$\pm$0.09~(14) & 8.03$\pm$0.12~~~(8) & 8.06$\pm$0.03~~~(2) & \nodata\\
$\varepsilon$(Mg\,{\sc ii})$^{\rm g}$ &  7.62$\pm$0.03~~~(3) & 7.60$\pm$0.01~~~(2)  & 7.58$\pm$0.10~~~(6)  & 7.56$\pm$0.03~~~(3)  & 7.51$\pm$0.10~~~(6)  & 7.50$\pm$0.05~~~(4) \\
$\varepsilon$(Si\,{\sc ii})$^{\rm h}$ &              \nodata &              \nodata &  7.47$\pm$0.17~~~(2) & 7.56$\pm$0.08~~~(2) & 7.51$\pm$0.10~~~(5) & 7.22$\pm$0.13~~~(6) \\
$\varepsilon$(Si\,{\sc iii})$^{\rm h}$&  7.50$\pm$0.08~~~(8) &  7.48$\pm$0.08~~~(6) &   7.46$\pm$0.11~~~(9) & 7.52$\pm$0.11~~~(8) & 7.53$\pm$0.17~~~(7) &7.29$\pm$0.05~~~(9) \\
$\varepsilon$(Si\,{\sc iv})$^{\rm h}$ &   7.50$\pm$0.04~(10) & 7.51$\pm$0.18~~\,(5) & 7.50$\pm$0.08~~\,(3) & 7.48$\pm$0.14~~\,(2) & 7.50$\pm$0.04~~\,(2)  & \nodata \\
$\varepsilon$(Fe\,{\sc ii})$^{\rm i}$ &              \nodata &              \nodata &              \nodata &              \nodata &          7.38~~~(1) &          7.38~~~(1) \\
$\varepsilon$(Fe\,{\sc iii})$^{\rm j}$&   7.38$\pm$0.12~(17) &  7.49$\pm$0.12~~~(5) &   7.44$\pm$0.09~(33) &   7.48$\pm$0.10~(30) &   7.42$\pm$0.12~(36) & 7.40$\pm$0.09~(32) \\
\tableline
\end{tabular}
\end{center}
\vspace{-3mm}
$\varepsilon({\rm El})$\,$=$\,$\log({\rm El}/{\rm H})$\,$+$\,12, with rms uncertainties and number of analysed lines in parentheses.
NLTE model atoms: H: \citet{przybilla04}; 
$^{\rm a}$\,\citet{przybilla05};
$^{\rm b}$\,\citet{nieva06,nieva08}; $^{\rm c}$\,\citet{przybilla01}; $^{\rm
d}$\,\citet{przybillaetal00}; $^{\rm e}$\,\citet{becker88},updated; $^{\rm
f}$\,\citet{morel08}, $gf$-values of \citet{fft04} for \ion{Ne}{1}; $^{\rm g}$\,\citet{przybillaetal01}; $^{\rm
h}$\,\citet{becker90}, extended \& updated; $^{\rm i}$\,\citet{becker98};
$^{\rm j}$\,\citet{morel07}\\
\end{table*}}

The quantitative analysis of the sample stars was carried
out following the hybrid NLTE approach discussed by
\citet[][NP07]{nieva07} and NP08.
In brief, line-blanketed LTE model atmospheres were computed with ATLAS9
\citep{kurucz93} and NLTE line-formation calculations were performed using updated 
versions of our codes {\sc Detail} and {\sc Surface}. 
State-of-the-art model atoms were adopted (see Table~\ref{params}), which
allow atmospheric parameters and 
elemental abundances to be obtained with high accuracy.

\begin{figure}[b]
\plotone{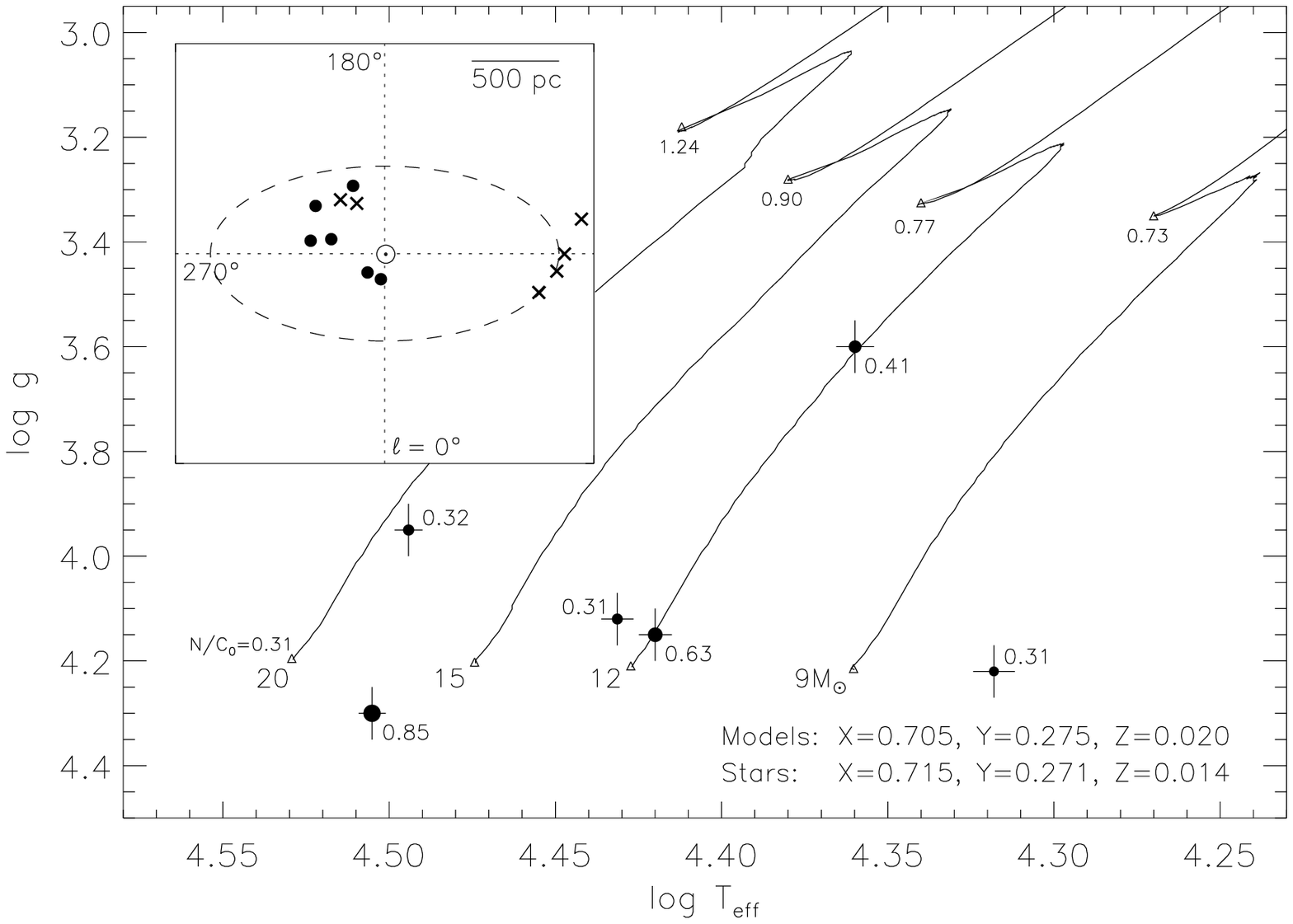}
\caption{Comparison with stellar evolution tracks \citep{meynet03}. Observed
and predicted N/C ratios (by mass) are indicated. The inset shows the
location of the sample stars as projected on the Galactic plane, with the
solar neighbourhood (as considered here) schematically outlined. 
Crosses denote positions of BA-SGs from the control sample (see text).
\label{evol}}
\end{figure}

Multiple independent spectroscopic indicators were considered simultaneously
for the determination of the atmospheric parameters, effective temperature
$T_{\rm eff}$ and 
surface gravity $\log g$: all Stark-broadened Balmer lines
and 4--6 ionization equilibria, of \ion{He}{1/II}, 
\ion{C}{2/III/IV}, \ion{O}{1/II}, \ion{Ne}{1/II}, \ion{Si}{2/III/IV} 
and \ion{Fe}{2/III}. Also, the observed spectral energy distributions 
were reproduced \citep{nieva06}. The 
redundancy
helps to avoid systematic errors. The microturbulent velocity $\xi$ was determined
by demanding that abundances be independent of line
equivalent widths. Elemental abundances $\varepsilon$(El),  
rotational velocity $v \sin i$ and macroturbulence $\zeta$ were determined from
fits to individual line profiles. The results are summarized in
Table~\ref{params}. Stellar parameters and He and C abundances 
are identical with those derived by NP07/NP08,
except for HR\,5285, where consideration of additional ionization equilibria
indicated small revisions, though agreement is obtained
within the mutual uncertainties. Spectral types 
and spectroscopic distances are also given in Table~\ref{params}, 
agreeing well with HIPPARCOS parallaxes (HR\,6165, HR\,5285)
or with the association distance (HR\,1861).
The positions of the stars in the $T_{\rm
eff}$--$\log g$-plane are indicated in Fig.~\ref{evol}, where a comparison with
evolution tracks 
is made. An overview of the location of the stars
in the solar vicinity is also given there.

The uncertainties in the atmospheric parameters were determined from the
quality of the simultaneous fits to all diagnostic indicators. Statistical
uncertainties for abundances were obtained from the
individual line data (rms values). Systematic errors in the abundances due
to uncertainties in atmospheric parameters, atomic data and the quality of the
spectrum are $\sim$0.1\,dex \citep[NP08,][]{przybillaetal06}, i.e. about as large
as the statistical errors.

\begin{figure*}
\centering
\includegraphics[width=.97\linewidth]{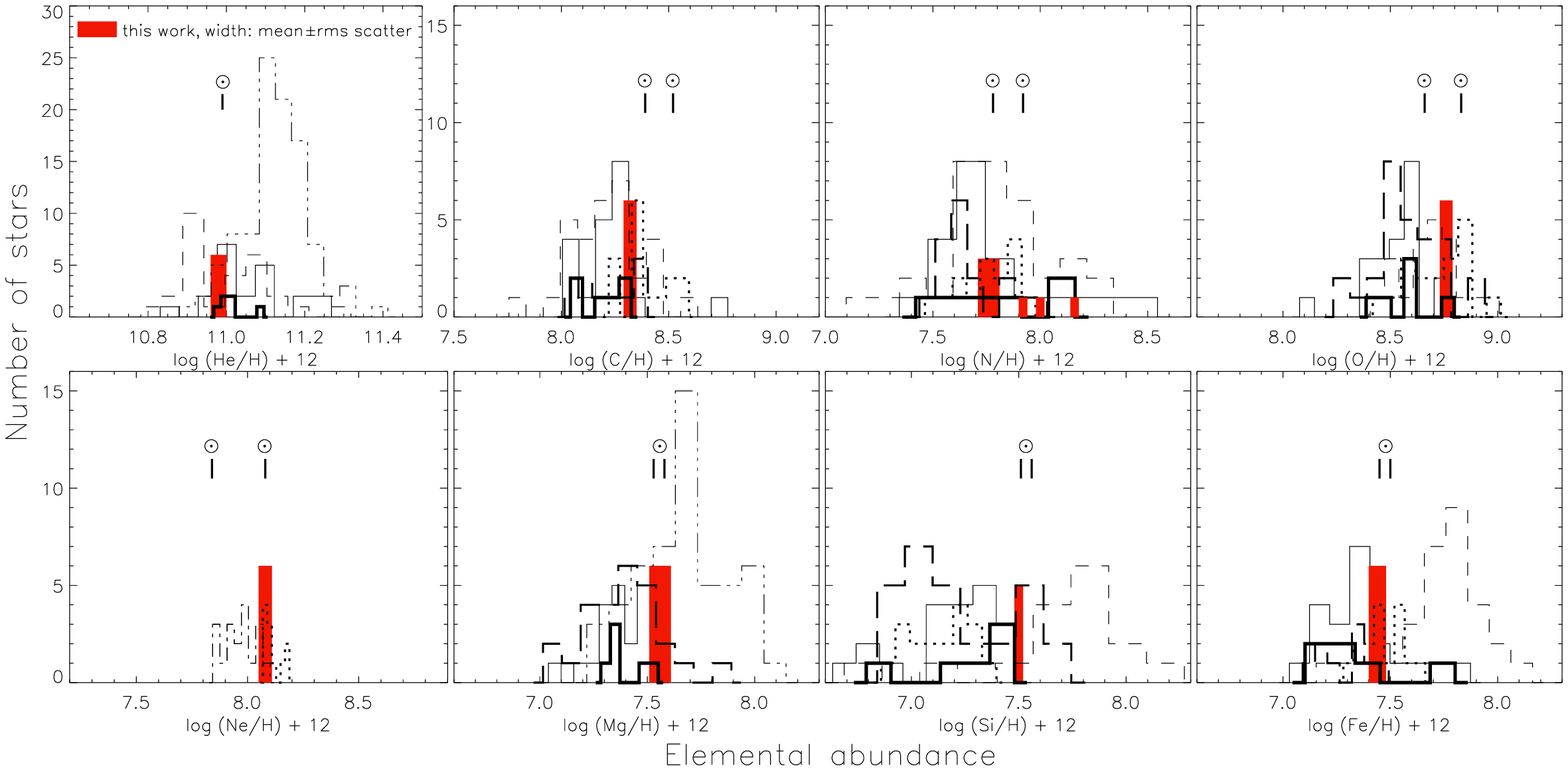}
\caption{Comparison of chemical abundance studies (NLTE) of B-stars in the
solar neighbourhood.
Red bars: present work; full line and thick full line (for the same stars
as in our work): \citet{kilian92,kilian94}; dotted: \citet{cunha94},
\citet{cunha06} for Ne;
short-dashed: \citet{gies92}, excluding bright giants and SGs;
long-dashed: \citet{daflon99,daflon01a,daflon01b,daflon03}; dot-dashed:
\citet{morel08};
triple-dot-dashed: \citet{lyubimkov04,lyubimkov05}. Bin width is $\sigma$/2
of the individual studies. Iron abundances in LTE in all
previous work. HR\,5285 is Si-peculiar and therefore excluded from the 
silicon mean abundance. Solar abundances ($\odot$): GS98 \& AGS05 (lower values).
The panel for C is reproduced from NP08 for completeness. 
See the text for details.\\ \label{histograms}}
\end{figure*}

Our analysis of each {\em individual} star differs from standard studies
in two main respects: {\sc i}) practically all (un\-blen\-ded) lines of
the ion spectra are analyzed instead of a few selected `good' lines
thus avoiding selection effects, and
{\sc ii}) all parameter indicators (in particular the
ionization equilibria) are closely matched simultaneously,
which has never been achieved before. As a result, practically the {\em entire} observed
stellar spectrum is reproduced closely by the spectral synthesis, 
see NP07, NP08, \citet{przybillaetal08} for examples.
This is facilitated by the use of critically evaluated data in the model
atom construction and a (time-consuming) iterative approach for a precise determination of
the stellar parameters 
(NP08). 
Less accurate photometric $T_{\rm eff}$-estimates as adopted in most previous work 
are avoided, as
$T_{\rm eff}$-uncertainties are often the most important sources of systematic 
error in the abundance derivation, next to ill-chosen atomic data (NP08) and
$\log g$-uncertainties (NP07).


\section{Chemical homogeneity of the solar vicinity}
The status of previous NLTE abundance studies of early B-stars in the solar 
neighbourhood (covering clusters, associations and field stars)
is illustrated in Fig.~\ref{histograms}. A wide range of abundance values
is found for most elements, typically spanning $\sim$1\,dex 
\citep[for comparison, such a range is bridged by
the cumulative effect of $\sim$13\,Gyrs of Galactochemical 
evolution, see e.g. Fig.~2 of][]{chiappini03}. Moreover,
the abundance distributions peak in most cases at sub-solar values, in particular 
when referring to 
the solar composition of GS98.
Exceptions are He, where most previous studies find values on average larger than
solar, and Ne (about solar, GS98, from two very recent studies; see also
Lanz et al. 2008 for the case of Ar). 
Several of these older B-star studies were combined by \citet{snow96} and
Sofia \& Meyer (2001, SM01, see Table~\ref{abus}) to derive a reference composition,
inevitably resulting in sub-solar average values and a large rms scatter.
The former
discrepancy has since been largely removed from a re-evaluation of solar abundances
(AGS05). However, the status quo in
terms of Galactochemical evolution can only be understood by invoking and fine-tuning extra
processes such as infall/outflow of material and local retention of supernova
products by large~amounts. 

On the other hand, our sample of early B-stars implies a high degree of homogeneity
for elemental abundances in the solar neighbourhood, with a scatter of $\sim$10\%, 
and absolute values of about solar (GS98 and/or AGS05, see Fig.~\ref{histograms} and 
Table~\ref{abus}). The only
exception is N, which is most sensitive to mixing of the atmospheric
layers with CN-processed material 
\citep[e.g.][]{maeder00}. In this case the
pristine N abundance may be indicated by the 3 objects with the lowest
value, implying a pristine N/C ratio of 0.31$\pm$0.05 (by mass; error bar
accounts for additional uncertainties).

Although our sample is small, we regard it as representative for the early 
B-star population in the solar neighbourhood. 
The stars sample the relevant portion of the H-burning phase of the objects in
the HRD in terms of $T_{\rm eff}$ and $\log g$ (see Fig.~\ref{evol}). They also sample 
one hemisphere of the solar neighbourhood (inset of Fig.~\ref{evol}), half of them 
located in OB associations and the other half in the field. 
All six stars were analyzed by \citet{kilian92,kilian94} before,
typically spanning the entire abundance range in her sample of 21 stars (see
Fig.~\ref{histograms}). We therefore also find a chance
selection of stars with similar chemical composition for our sample unlikely.
The wide abundance ranges found in previous work reflect the lower
accuracy of the analyses, while shifts of the abundance distributions
relative to each other reflect systematics, with different temperature
scales being the most important among these.
Our results are supported further by a control sample of six BA-type
supergiants (BA-SGs, Fig.~\ref{evol}), for which mean
values of $\varepsilon$(O)\,=\,8.80$\pm$0.02 and
$\varepsilon$(Mg)\,=\,7.55$\pm$0.07
were derived using the same analysis methodology as applied
here \citep{przybillaetal06,firnstein06}.

The finding of chemical homogeneity for our sample is in
excellent accordance with results from the analysis of the ISM gas-phase in
the solar neighbourhood \citep[][and references therein]{sofia04} and with theory regarding the
efficiency of hydrodynamic mixing in the ISM \citep[][]{edmunds75,roy95}.
Excellent agreement is also obtained with elemental abundances in the Orion
nebula \citep[][E04, see Table~\ref{abus}]{esteban04}, with the exception of
C, which may be a consequence of the atomic data used in the Orion analysis
(see NP08 for the stellar case) plus overestimated dust corrections.

In the following we briefly investigate the impact of this {\em cosmic abundance standard}
on important topics of contemporary astrophysics.

{
\begin{table*}
\footnotesize
\caption{Chemical composition of different object classes in the solar neighbourhood and of the Sun\label{abus}}
\begin{center}
\vspace{-5mm}
\begin{tabular}{lr@{/}r@{~~~~~~~}rrrrrr@{/}r}
\tableline\\[-2.5mm]
\tableline
        & \multicolumn{2}{c}{cosmic standard}     & Orion              &                   & Young                &  ISM                     & ISM\\
Elem.   & \multicolumn{2}{c}{B stars -- this work}& gas+dust$^{\rm b}$  & B stars$^{\rm c}$ & F\&G stars$^{\rm c}$     &  gas                     & dust$^{\rm d}$ & \multicolumn{2}{c}{Sun$^{\rm e/f}$}\\
\tableline                                                                                                             
He      & 10.98$\pm$0.02 & \nodata$^{\rm a}$       &    10.988$\pm$0.003 &           \nodata &              \nodata &  \nodata                 & \nodata        & \multicolumn{2}{c}{10.99$\pm$0.02}\\
C       &  8.32$\pm$0.03 & 209$\pm$15             &      8.52$\pm$0.02  &     8.28$\pm$0.17 &        8.55$\pm$0.10 &  8.15$\pm$0.06$^{\rm g}$ & 68$\pm$26      & 8.52$\pm$0.06 & 8.39$\pm$0.05\\
N       &  7.76$\pm$0.05 &  58$\pm$\phantom{4}7   &      7.73$\pm$0.09  &     7.81$\pm$0.21 &              \nodata &  7.79$\pm$0.03$^{\rm h}$ & \nodata        & 7.92$\pm$0.06 & 7.78$\pm$0.06\\
O       &  8.76$\pm$0.03 & 575$\pm$41             &      8.73$\pm$0.03  &     8.54$\pm$0.16 &        8.65$\pm$0.15 &  8.59$\pm$0.01$^{\rm i}$ & 186$\pm$42     & 8.83$\pm$0.06 & 8.66$\pm$0.05\\
Ne      &  8.08$\pm$0.03 & 120$\pm$\phantom{4}9   &      8.05$\pm$0.07  &           \nodata &              \nodata &  \nodata                 & \nodata        & 8.08$\pm$0.06 & 7.84$\pm$0.06\\
Mg      &  7.56$\pm$0.05 &  36$\pm$\phantom{4}4   &            \nodata  &     7.36$\pm$0.13 &        7.63$\pm$0.17 &  6.17$\pm$0.02$^{\rm j}$ & 34.8$\pm$4.4   & 7.58$\pm$0.05 & 7.53$\pm$0.09\\
Si      &  7.50$\pm$0.02 &  32$\pm$\phantom{4}1   &            \nodata  &     7.27$\pm$0.20 &        7.60$\pm$0.14 &  6.35$\pm$0.05$^{\rm j}$ & 29.6$\pm$2.2   & 7.55$\pm$0.05 & 7.51$\pm$0.04\\
Fe      &  7.44$\pm$0.04 &  28$\pm$\phantom{4}3   &            \nodata  &     7.45$\pm$0.26 &        7.45$\pm$0.12 &  5.41$\pm$0.04$^{\rm j}$ & 27.3$\pm$2.7   & 7.50$\pm$0.05 & 7.45$\pm$0.05\\
\tableline
\end{tabular}
\end{center}
\vspace{-3mm}
$^{\rm a}$\,in units of $\log ({\rm El}/{\rm H})+12$\,/\,atoms per 10$^6$ H
nuclei -- computed from average star abundances (mean values over all individual 
lines {\em per element}, equal weight per line);
$^{\rm b}$\,E04; 
$^{\rm c}$\,SM01; 
$^{\rm d}$\,difference between the cosmic standard 
and ISM gas-phase abundances, in units of atoms per 10$^6$ H nuclei;
$^{\rm e/f}$\,GS98/AGS05, photospheric values; 
$^{\rm g}$\,\citet{sofia04};
$^{\rm h}$\,\citet{meyer97}, corrected accordingly to \citet{jensen07};
$^{\rm i}$\,\citet{cartledge04};
$^{\rm j}$\,\citet{cartledge06}\\
\end{table*}}


\section{The cosmic abundance standard: first applications}  
In general, excellent agreement of our B-star abundances with solar values from recent
3D radiative-hydrodynamical simulations of the solar atmosphere
(AGS05) is obtained. The oxygen value falls between GS98 and AGS05 values
and neon is compatible with GS98.
This opens up new perspectives in the ongoing discussion on helioseismic constraints,
chemical abundances and the solar interior model as reviewed by \citet{basu08}.

Our cosmic abundance standard also facilitates a precise determination of
dust depletion in the local ISM for the primary constituents. 
The amount of material incorporated into
dust grains is determined by the difference between our B-star abundances and
the ISM gas-phase abundances, 
see Table~\ref{abus}. Accordingly, a composition poor in carbon but rich in oxygen
and refractory elements is indicated.

Such studies were undertaken previously,
using e.g.\ abundances of the Sun, of B stars and of young F \& G stars
\citep[e.g.][SM01, see Table~\ref{abus}]{snow96} as proxies for the determination
of the dust-phase composition, however with mixed success. 
In particular, B stars were rejected as reliable indicators as the derived
abundances of material in dust at that time were too low to produce the 
observed interstellar extinction. Our study revives B stars as proxies of
the ISM dust-phase composition, and even more so because of the extremely
low abundance scatter compared to all other standards considered so far,
except for the Sun. 

The present results imply tight observational constraints on dust models in terms of
carbon abundance. The observed properties of dust grains, as
inferred from the interstellar extinction law, have to be produced by a 
rather small amount of carbon, posing a challenge to most dust models
\citep[see e.g.][]{snow95}. 
We can carry out an important consistency check, following
\citet{cartledge06}:
the O predicted to be incorporated in grains
from the observed Mg, Si and Fe dust abundances and a
rudimentary dust model agrees with the derived O dust abundance within 
the mutual (small) uncertainties. For the rudimentary dust model we assume 
silicates to be predominantly MgSiO$_3$, with only a small fraction of Fe
bound in silicates and only a small fraction being of olivine-like
composition. The remaining Mg and Fe fraction~is considered to be
in oxide form (MgO, FeO, Fe$_2$O$_3$, Fe$_3$O$_4$), see e.g. \citet{draine03} for a
discussion of observational evidence. 

Finally, we combine our B-star abundances with data for S, Cl and Ar
from the analysis of the Orion nebula (E04) and solar meteoritic
values for other abundant refractory elements (with
$\varepsilon$(El)\,$\gtrsim$\,5, AGS05)
to derive mass fractions for hydrogen, helium and the metals. 
Values of $X$\,$=$\,0.715, $Y$\,$=$\,0.271, $Z$\,$=$\,0.014 and
$Z/X$\,$=$\,0.020 characterize the present-day cosmic matter in the
solar neighbourhood \citep[to be compared to protosolar values
$X_0$\,=\,0.7133, $Y_0$\,=\,0.2735 and $Z_0$=0.0132,][]{GAS07}. 
These combined abundances are our recommended values
for a wide range of applications requiring an accurate knowledge of the
chemical composition at present (e.g.\ for opacity calculations), examples
being models of star/planet formation or stellar evolution (in particular of
short-lived massive stars), or for the empirical calibration of
Galactochemical evolution models.

\acknowledgments
We express our deep gratitude to U. Heber for support and useful 
comments on the manuscript, and thank 
M. Asplund and A. Serenelli 
for stimulating discussion.
M.F.N. acknowledges support by DFG (grant HE\,1356/45-1).


\end{document}